\documentclass[12pt]{article}
\title{Momentum space topology of fermion zero modes on brane}
\author{G.E. Volovik
\\Low Temperature Laboratory, Helsinki University of
Technology\\
P.O.Box 2200, FIN-02015 HUT, Finland\\
\\
 L.D. Landau Institute for
Theoretical Physics\\  Kosygin Str. 2, 117940 Moscow, Russia
}

\begin{document}
\maketitle

\begin{abstract}
{We discuss fermion zero modes within the 3+1 brane -- the domain wall
between the
two vacua in 4+1 spacetime. We do not assume relativistic invariance
in 4+1
spacetime, or any special form of the 4+1 action. The only input is
that the fermions in bulk are fully gapped and are described by
nontrivial
momentum-space topology. Then the 3+1 wall between such vacua contains
chiral 3+1 fermions. The bosonic collective modes in the wall form the
gauge and
gravitational fields. In principle, this universality class of
fermionic
vacua can
contain all the ingredients of the Standard Model and gravity.
 }
\end{abstract}

{\it Introduction.} The idea that our Universe lives on a brane
embedded in
higher dimensional space \cite{RubakovShaposhnikov1983} is popular at
the
moment. It is the further development of an old ideas of extra compact
dimensions introduced by Kaluza \cite{Kaluza} and Klein \cite{Klein}.
In a
new approach the compactification occurs because the low-energy
physics is
concentrated within the brane, for example, in a flat 4-dimensional
brane
embedded in a 5-dimensional anti-de Sitter space with a negative
cosmological
constant \cite{RandallSundrum}. Branes can be represented by
topological
defects, such as domain walls (membranes) and strings. It is supposed
that
we live inside the core of such a defect. This new twist in the idea
of extra
dimensions is
fashionable because by accomodation of the core size one can bring the
gravitational Planck energy scale close to TeV range. That is why,
there is
a hope
that the deviations from the Newton's law can become observable
already at the
distance of order 1 mm. At the moment the Newton's law has been
tested for
distances $>0.2$ mm \cite{experiment}.

The particular mechanism of why the matter is localized on
the brane, is that the low-energy fermionic matter is represented by
the
fermion zero modes, whose wave function is concentrated in the core
region.
Outside the core the fermions are massive and thus are frozen out at
low
temperature $T$. An example of such topologically induced Kaluza-Klein
compactification of the multidimensional space is provided by the
condensed
matter analogs of branes -- domain walls and vortices. These
topological
defects do contain fermion zero modes which can live only within the
core of
defects. These fermions form the 2+1 world within the domain wall,
and 1+1
world within the domain wall in quasi-two-dimensional thin films or
in the
core of the linear defects -- quantized vortices.

Recently an attempt was made to `construct' the 4+1 condensed matter
system
with
gapped fermions in bulk and gapless excitations on the 3+1 boundary,
which
include gauge bosons and gravitons \cite{Zhang4+1Hall}. This is the
4+1
dimensional generalization of the 2+1 quantum Hall effect (QHE) in the
presence of the external $SU(2)$ gauge field, where the low energy
fermions
are the analogs of the so-called edge states \cite{EdgeStates} on the
boundary of the system.

Here we show that there is a natural scenario in which the
chiral fermions emergently appear in the brane together with gauge and
gravitational field.  Instead of the QHE system, we consider the
system in
which quantization of Hall conductivity occurs without external
magnetic
field. In this scenario the topology of momentum space
\cite{Volovikreview,MomentumSpaceTopologyReview} plays the central
role
determining the universality classes. We consider the domain wall,
which
separates two 4+1 quantum vacua with nontrivial topology in the
momentum
space. If the momentum-space topological invariant is different on
two sides
of the wall, such 3+1 brane contains fermion zero modes -- the
gapless 3+1
fermions. Close to the nodes in the energy spectrum -- Fermi points --
 these
fermions
are chiral. The collective bosonic modes within the brane correspond
to
gauge and gravitational fields acting on fermion modes.

As distinct from the
relativistic theories
\cite{RandallSundrum}, in this scenario the
existence of gauge and gravitational fields in brane does not require
the
existence of the corresponding 4+1 fields in the bulk.
The 3+1 fields in the brane emergently arise as collective modes of
fermionic
vacuum in the same manner as they arise in quantum liquids belonging
to the
universality class of Fermi points
\cite{Volovikreview}. Thus, the brane separating the 4+1 vacua with
different
momentum-space topology is one more universality class of the quantum
vacua, whose
properties are dictated by the momentum-space topology.

{\it Walls in 2+1 systems.} Let us first consider how all this occurs
in 2+1
systems, then it can be easily generalized to the 4+1 case. For the
2+1
systems it is known that the quantization of Hall or spin-Hall
conductivity
can occur even without an external magnetic field. This quantization
is
provided by the integer valued momentum-space topological invariant
\cite{VolovikYakovenko}:  \begin{equation} N_3 =
{1\over{24\pi^2}}e_{\mu\nu\lambda}~ {\bf tr}\int   dp_xdp_ydp_0 ~
{\cal
G}\partial_{p_\mu} {\cal G}^{-1} {\cal G}\partial_{p_\nu} {\cal G}^{-
1} {\cal
G}\partial_{p_\lambda}  {\cal G}^{-1}~.  \label{3DTopInvariant}
\end{equation}
Here ${\cal G}$ is the fermionic propagator expressed in terms of the
momentum $p_\mu=({\bf p}, p_0)$, where ${\bf p}=(p_x,p_y)$ and $p_0$
is the
frequency on the imaginary axis. In the most simple examples, which
occur for
example in thin films of $^3$He and probably in the atomic layers of
some
superconductors, one has ${\cal G}^{-1}=z -{\cal H}({\bf p})$, where
$z=ip_0$; and the $2\times 2$ Hamiltonian ${\cal H}({\bf p})= \tau^i
 g_i(p_x,p_y)$ is expressed in terms of Pauli matrices $\tau^i$. In
this case
the Eq.(\ref{3DTopInvariant}) is simplified:  \begin{equation} N_3=
{1\over
4\pi}\int {dp_xdp_y\over |{\bf g}|^3}~{\bf g}\cdot \left({\partial
{\bf
g}\over\partial {p_x}} \times {\partial {\bf g}\over\partial
{p_y}}\right)~.
\label{2DInvariant}
\end{equation}
The invariant exists only if the fermions are gapped, i.e. their
energy $E(p_x,p_y)=|{\bf g}|\neq 0$. The value of Hall or spin-Hall
conductivity
depends on this invariant and that is why the quantization of
conductivities
occurs without external field. The invariant $N_3$ can be varied by
varying the
thickness of the film, instead of varying the magnetic field in
conventional QHE.
Similar invariants have been used in conventional QHE too, see
\cite{Kohmoto,Ishikawa}.

Another important property of the conventional QHE, which is
reproduced by the
system under discussion, is the existence of the edge states on the
boundary of the
system, or on the boundary separating vacua with different values of
quantized
conductivity. Let us consider the domain wall (the 1+1 brane)
separating vacua
with different topological invariants on the left and on the right
side of the
wall: $N_3(right)$ and $N_3(left)$. If $N_3(right)\neq N_3(left)$ one
finds
that there are fermion zero modes.
These are
the gapless branches $E(p_\parallel)$, where
$p_\parallel$ is the linear momentum along the wall. These branches
cross zero
energy when $p_\parallel$ varies.
Close to zero energy the spectrum of the $a$-th fermion zero mode is
linear:
\begin{equation}
E_a(p_\parallel)=c_a (p_\parallel -p_a)~~
\label{FermionZeroModesSoliton1}
\end{equation}
These fermion zero modes correspond  to
the chiral (left-moving and right-moving) gapless edge states in QHE.
There is an index theorem which determines the  algebraic number
$\nu$  of the fermion zero modes, i.e. the number of modes crossing
zero with
positive slope (right-moving) minus the number of modes with negative
slope (left-moving):
\begin{equation}
\nu=\sum_a {\rm sign}~c_a ~~.
\label{FZMSolitonIndexTheorem}
\end{equation}
According to this theorem, which is similar to the Atiyah-Singer
index theorem
\cite{AtiyahSingerIndexTheorem} relating the number of fermion zero
modes with the topological charge of the gauge field
configuration, one has \cite{VolovikEdgeStates}
\begin{equation}
\nu=  N_3(right)- N_3(left) ~~.
\label{FermionZeroModesSoliton3}
\end{equation}
The crossing point $p_a$ on each branch is nothing but the Fermi
surface in
1D momentum space $p_\parallel$. In general the Fermi surfaces
can be described by the topological invariant $N_1$
expressed in terms of Green's function \cite{Volovikreview}
\begin{equation}
N_1={\bf Tr}~\oint_C {dl\over 2\pi i} {\cal
G}(p_0,p_\parallel)\partial_l
{\cal G}^{-1}(p_0,p_\parallel)=\nu~.  \label{InvariantForEdgeStateFS}
\end{equation}
Here ${\cal G}$ is the propagator for the 1+1 fermion zero modes; and
the contour $C$ embraces all the points $(p_0=0,p_\parallel=p_a)$
where the
Green's function is singular.  In the simplest case the propagator
for the
1+1 fermion zero modes has the form \begin{equation} {\cal G}^{-1}=
ip_0 -
E_a(p_\parallel)~, \label{GFerZeorModesVierbeinWall} \end{equation}
and the
contour $C$ embraces the point $(p_0=0,p_\parallel=p_a)$ in momentum
space.
The equation $N_1= N_3(right)- N_3(left)$ illustrates the topology of
the
dimensional reduction in the momentum space: the momentum-space
topological
invariant  $N_3$ of the bulk 2+1 system gives rise to the  1+1
fermion zero
modes
described by the momentum-space topological invariant $N_1$.

{\it Branes in 4+1 systems.} Now we can move from 2+1 to 4+1
dimension.
Let us suppose that we have the quantum liquid in 4+1 spacetime,
which contains
the 3+1 domain wall separating two domains, each with fully gapped
fermions. Then everything can be obtained from the case of the quantum
liquid in
2+1 spacetime just by increasing the dimension.

According to analogy with 2+1 systems, the 4+1 gapped fermions must
have nontrivial momentum-space topology. Such topology is described
by the
invariant
$N_5$ instead of $N_3$:
\begin{equation}
N_5 = C_5
 e_{\mu\nu\lambda\alpha\beta}~{\bf tr}~\int~  d^5p
~ {\cal G}\partial_{p_\mu} {\cal G}^{-1}
{\cal G}\partial_{p_\nu} {\cal G}^{-1} {\cal G}\partial_{p_\lambda}
{\cal
G}^{-1} {\cal G}\partial_{p_\alpha} {\cal G}^{-1}{\cal
G}\partial_{p_\beta}
{\cal G}^{-1}~.
\label{TopInvariantMatrixTildeN5}
\end{equation}
Here $p_\mu=(p_0,{\bf p})$, where  ${\bf p}=(p_1,p_2,p_3,p_4)$ is the
momentum in
4D space; $p_0$ is the energy considered at imaginary axis: $z=ip_0$;
and $C_5$
is proper normalization. It is the difference $N_5(right)-
N_5(left)$ of invariants on both sides of the domain wall (extension
of
$N_3(right)-  N_3(left)$), which must give rise to the 3+1 fermion
zero
modes within the brane.

The relativistic example of the propagator
with nontrival invariant $N_5$ is provided by ${\cal G}^{-1}=ip_0 -
{\cal H}$,
where the Hamiltonian in the 4D space is ${\cal
H}=M\Gamma^5+\sum_{i=1}^4\Gamma^i p_i$, and $\Gamma^{1-5}$ are
$4\times 4$
Dirac matrices satisfying the Clifford algebra
$\{\Gamma^{a},\Gamma^{b}\}=2\delta^{ab}$. In this example the proper
domain
wall,
which contains the fermion zero modes, separates the domains with
opposite
signs
of the mass parameter
$M$, since for such wall $ N_5(right)=-
N_5(left)$. The existence of fermion zero modes in such domain wall
 is a well known fact in relativistic theories. We would like to
stress,
however, that the existence of fermion zero modes does not require the
relativistic
theory in the bulk. It is enough to have the nontrivial invariant
$N_5$, which
determines the universality class of fermionic vacuum.

In the 1+1 wall the energy spectrum of fermion zero modes in the wall
crosses
zero at points in 1D momentum space. Thus the energy spectrum of
fermion zero
modes in
the 3+1 brane must be zero at points in 3D momentum space. This means
that the
spectrum has Fermi points. Fermi points are described by the momentum-
space
topological invariant
$N_3$, which is now the difference between the number of right-handed
and
left-handed fermions \cite{Volovikreview}:
\begin{equation} N_3 =
{1\over{24\pi^2}}e_{\mu\nu\lambda\gamma}~ {\bf tr}\int_{\sigma_3}~
dS^{\gamma} ~ {\cal G}\partial_{p_\mu} {\cal G}^{-1} {\cal
G}\partial_{p_\nu}
{\cal G}^{-1} {\cal G}\partial_{p_\lambda}  {\cal G}^{-1}~.
\label{TopInvariant}
\end{equation}
Here the integral is over
the 3-dimensional surface
$\sigma_3$ embracing the singular points $(p_0=0,{\bf p}={\bf
p}_a)$ of the spectrum. This is the analog of invariant
$N_1$ in Eq.(\ref{InvariantForEdgeStateFS}).
 Close to the $a$-th
Fermi point the fermion zero modes represent 3+1 chiral fermions,
whose
propagator
has the general form expressed in terms of the tetrad
field:
\begin{equation}
{\cal G}^{-1}=\sigma^\nu e^\mu_{\nu a} (p_{\mu} - p_{\mu a})
~.
\label{GeneralPropagator}
\end{equation}
Here $\sigma^\nu =(1,{\bf\sigma})$, and ${\bf\sigma}$ are Pauli
matrices.

In the same manner as in Eqs.(\ref{FermionZeroModesSoliton3}) and
(\ref{InvariantForEdgeStateFS}) which relate the number of fermion
zero modes
with the topological invariants in bulk 2+1 domains, the total
topological
charge
of the Fermi points within the domain wall is expressed through the
difference of
the topological invariants in bulk 4+1 domains:
\begin{equation}
N_3=N_5(right)-  N_5(left)~.
\label{NumberOfZeroModes}
\end{equation}

The quantites $e^\mu_{\nu a}$ and $p_{\mu a}$, which enter the
fermionic
spectrum,
are dynamical variables. These are the low energy collective bosonic
modes
which play the part of the effective gravitational and gauge fields
correspondingly
acting on chiral fermion \cite{Volovikreview}. These fields emergently
arise in the
fermionic vacuum with nontrivial momentum-space topology. The brane
between the
topologically different vacua thus represents one more universality
class
of the
`emergent behavior'
\cite{LaughlinPines}.

In a similar manner the gauge and gravity fields arise as collective
modes on the boundary of the 4+1 system exhibiting the quantum Hall
effect
\cite{Zhang4+1Hall}. Both systems have similar topology: in
Ref.\cite{Zhang4+1Hall} the nontrivial topology is provided by the
external
field, while in our case it is assumed that the vacuum itself has
a nontrivial topology, $N_5\neq 0$, even without the gauge field.

{\it Conclusion.} We showed that if the momentum-space topology of the
fermionic vacuum in 4+1 spacetime is nontrivial, the 3+1 domain wall
between
the two such vacua contains chiral fermions, while bosonic collective
modes
in the wall are the gauge and gravitational fields. This emergent
behavior
does not depend on details of the action in the bulk 4+1 systems, or
on
details of the structure of the brane. Neither Lorentz invariance,
nor the
gravity in 4+1 bulk system are required for emergency of chiral
fermions and
collective fields in the brane.  However, the nontrivial topology
alone does
not guarantee that the gravitational field will obey Einstein
equations: the
proper (maybe discrete) symmetry and the proper relations between
different
"Planck" scales in the underlying fermionic system are required
\cite{Volovikreview}.  The energy scale which marks the cut-off of the
integrals over fermions must be much smaller than the energy scale at
which
the Lorentz invariance is violated. We hope that within this
universality
class one can obtain all the ingredients of the Standard Model and
gravity.

I thank V.A. Rubakov and S. Zhang for fruitful discussions.
This work was supported by ESF COSLAB Programme and by the Russian
Foundations
for Fundamental Research.

\end{document}